\documentclass[prb,preprint,aps,showpacs]{revtex4}
\usepackage{graphicx}
\begin{document}

\title{ Inter-grain tunneling  in the half-metallic double-perovskites
Sr$_2$BB'O$_6$ (BB'-- FeMo, FeRe, CrMo, CrW, CrRe)}

\author{B. Fisher}
\email{phr06bf@physics.technion.ac.il}
\author{J. Genossar}
\author{K. B. Chashka}
\author{ L. Patlagan}
\author{G. M. Reisner}
\affiliation{Physics Department, Technion, Haifa 32000, Israel. }
\date{\today}

\begin{abstract}
The zero-field conductivities ($\sigma$) of the polycrystaline
title materials, are governed by inter-grain transport. In the
majority of cases their $\sigma$(T)  can be described by the
"fluctuation induced tunneling"  model. Analysis of the
results in terms of this model reveals two remarkable features: 1.
For \emph{all} Sr$_2$FeMoO$_6$ samples of various microstructures,
the tunneling constant (barrier width $\times$ inverse
decay-length of the wave-function)  is $\sim$ 2, indicating the
existence of an intrinsic insulating boundary layer with a
well defined electronic (and magnetic) structure. 2. The tunneling
constant for \emph{all} cold-pressed  samples decreases linearly
with increasing magnetic-moment/formula-unit.
\end{abstract}
 \maketitle


Half-metallic ordered double-perovskites with fully polarized
conduction bands and Curie temperatures (T$_c$) above room
temperature (RT) are of interest for devices which depend on spin
polarized transport. Therefore their magnetic, electronic and in
particular their magneto-resistive properties{\cite{serrateJPCM}} have
been investigated intensively over the past two decades.
 The grain boundaries in these materials act in most cases as tunnel barriers.  The early theories of inter-grain tunneling magneto-resistance addressed the problem of tunneling through a non-magnetic barrier separating two ferromagnetic grains (including vacuum).{\cite{julliere,slonczewski,inoue}}  These theories could not explain inter-grain magneto-resistance in half metals. In Ref. {\onlinecite{serrate}}  the magneto-resistive behaviour of
 (BaSr)$_2$FeMoO$_6$ was explained  in terms of tunneling between two correlated spin glass-like surfaces separated by a thin insulating
layer. In Ref. {\onlinecite{jana}} it was suggested that the pinned ferromagnetic spins at the
core/skin interface should be taken as being solely responsible
for the tunneling magneto-resistance in half-metallic double-perovskites. A spin-glass-like surface layer surrounding each soft ferromagnetic (FM) grain of Sr$_2$FeMoO$_6$ has been detected also in Ref. {\onlinecite{ray}} by careful ac susceptibility measurements on
a highly ordered polycrystalline sample; these measurements were able to separate the barrier layer signal from the bulk. The presence of an intrinsic insulating boundary layer around FM grains of (LaSr)MnO$_3$ (LSMO),  with magnetic properties different from those of the bulk, has been recently revealed by means of x-ray linear dichroism and transport measurements.{\cite{arx}} This phase,  about 2 unit cells thick, is held responsible for the observed depressed magneto-transport properties in manganite based magnetic tunneling junctions.

Unlike the difficulty in
separating the magnetic properties of the layers from those
of the bulk,{\cite{ray}} it is relatively easy to study
the  electronic properties of the grain skin layers  when the electronic
transport  is dominated by inter-grain tunneling as is
the case in most of the polycrystalline samples of the
title materials. In this report we focus on the zero-field conductivity of various samples of the five title compounds; this comparative study
 revealed some important features of the  grain-boundaries of these half metals.

Table I  shows the five title double-perovskites  (with
abbreviations), their ionic configuration, nominal (ideal)
saturation magnetization (M$_i$) and T$_c$. While
their bulk is metallic, as confirmed by their metallic-like
thermopower, {\cite{thermopower}} the zero-field conductivities
($\sigma(T)$)  of polycrystalline samples are non-metallic (the
conductivity increases with increasing T). Metallic-like resistivity was found in a single crystal of SFMO.{\cite{tomioka}} The inter-grain
tunneling conductivity depends strongly on preparation conditions
and often exhibits unusual T-dependence. The most remarkable
behaviors are the linear-in-T  conductivities from liquid He temperatures up to RT, for
\textit{all} our sintered and granular samples of SFMO ,
irrespective of preparation conditions, for some samples of SFRO
and of SCMO,  and the
linear-in-T$^2$  conductivity over the same range of T, for some samples of
SCMO.{\cite{remarkable}} The
temperature dependence of the conductivity for all our samples,
except for porous SCRO, can be derived from the "fluctuation
induced tunneling" (FIT) model. {\cite{sheng}} This model applies
to metallic grains embedded in an insulating medium. Tunneling
occurs across small gaps (width $w$ and area A) between large
metallic grains; the small gaps are subject to large thermal
fluctuations of the voltage.

$\sigma$(T) predicted by this model is:
\begin{equation} \sigma=\sigma_o exp(- {{ T_1}\over {T_o+T}})=\sigma(0) exp({{T_1T}\over {T_o(T_o+T)}})
\label{eq:FIT}
\end{equation}
where $k_BT_1=(2/\pi)(A/w)(V_o/e)^2$ is the electrostatic energy
within a parabolic potential barrier of width $w$ and height $V_o$
of a junction of area $A$; $T_1/T_o =\pi\chi w/2$ is the tunneling
constant where $\chi={\sqrt{2mV_o/{\hbar}^2}}$, $\sigma_o$ is a pre-exponent that may be regarded as independent of temperature and $\sigma(0)=\sigma_o exp (-T_1/T_o)$ . The FIT equation for
$\sigma$(T) is an extension of the formula derived for a single junction
to a network of fluctuating tunneling junctions.{\cite{sheng}} For
$T\ll {T_o}$,  Eq. (\ref{eq:FIT}) represents elastic tunneling and for $T
\gg {T_o}$ - activated conductivity with activation energy
$k_BT_1$. The effect of the thermal fluctuations is to reduce the
barrier's height and width; for  $T=T_o$ the effective tunneling
constant is half its value at $T=0$. This equation includes the
rare and interesting cases mentioned above for specific ranges of
the parameter $T_1/T_o$ and of $T/T_o$. In Ref.
{\onlinecite{remarkable}} we showed that for $T_1/T_o <3$ a linear
function of $T$ fits $\sigma(T)$ over a $T/T_o$ range that
increases with $T_1/T_o$. The correlation parameter of the linear
fit to Eq. (\ref{eq:FIT}) for $T_1/T_o \lesssim 3$ and $T/T_o\leq 1.1$ is
$R^2=0.9999$. In this range, $\sigma(T)$ varies up to a factor of
5, in good agreement with our findings (see  Fig. 4 in Ref.
{\onlinecite{remarkable}}) . We showed also that ($\sigma(T)-
\sigma(0))\propto T^2$ $(R^2=0.9999)$, for a narrow range of
$T_1/T_o$ around 8 and $T/T_o$ up to $\sim 1.8$. Within this range
$\sigma(T)$ may vary by more than two orders of magnitude, again
in good agreement with our findings (see Fig. 3 in Ref.
{\onlinecite{remarkable}}).

All SFMO samples exhibit linear $\sigma(T)$. In
other compounds conductivity linear-in-$T$ or linear-in-$T^2$  (over a wide
temperature range) are special cases. However, except for SCRO,
$\sigma(T)$ for all sintered samples obeys the FIT model with parameters within a
wide range that depend on the preparation conditions.
Our sintered SCRO samples were porous and their conductivity  over an
unprecedentedly wide range of T  was
of Berthelot-type ($ln(\sigma(T)/\sigma(0)) = T/T_B$ where $T_B$
is a constant of the order of a few tens of K).{\cite{SCRO}} This behavior can
be derived from Tredgold's "vibrating barrier tunneling" model.
{\cite{tredgold}}  Eq. (\ref{eq:FIT}) can be
reduced to a Berthelot-type formula for $T_1/T_o >> 1$ and $T/T_o
<< 1$ with $T_B =T_o^2/T_1$ but then the values of the fitting
parameters to our data become non-physical. Interestingly,
$\sigma(T)$ for cold-pressed (c.p.) SCRO obeys the FIT model with
reasonable parameters (see below).

The FIT model has been extended to  electric-field dependent
conductivities.  The nonlinear I-V characteristics measured on
some SFMO samples (using pulsed currents in order to avoid Joule
heating) are consistent with the extended FIT model, at least
qualitatively .{\cite{efield}}

The FIT model does not address magnetic interactions. Since
it was applied successfully to at least three groups of
magnetic materials (our title materials, CrO$_2$ and its
composites{\cite{bajpal,fan}} and Co-based nanocomposites
{\cite{wen}}), it may be assumed that the influence of the magnetic interactions is on the nature of
the tunneling barrier and on the pre-exponent. Since $M_i$ of our samples varies between 1 and 4, we
attempted to detect correlations between the tunneling
parameters of the exponent of Eq. (\ref{eq:FIT}) and $M_i$.

Table II contains the fitting parameters of Eq. (\ref{eq:FIT}) to the experimental $\sigma(T)$: $\sigma(0)$, $T_1$, $T_o$ and
$T_1/T_o$,  for our
samples of Sr$_2$BB'O$_6$. The labels of the five groups of
samples are followed by the sources of the data (the relevant
references to our previous publications and  Figs. 1 and 2 shown
here), M$_i$ and  fitting
parameters. Only two parameters in the exponent are independent
but for convenience all three are shown ($T_1$, $T_o$ and
$T_1/T_o$). We also show the fitted
parameters for a cold-pressed Sr$_{1.5}$La$_{0.5}$FeMoO$_6$
(LSFMO).  Additional plots of $\sigma(T)$ for
SFMO are shown in Ref. {\onlinecite{efield}}; all are straight
lines up to RT. The slopes of the plots for the cold-pressed samples are
steeper than those for the sintered samples.
The conductivities of the samples at  T=0 ($\sigma(0)$)  spread over many orders of magnitude, from $10^{-6}$ to $10^2$ $(\Omega cm)^{-1}$. The highest $\sigma(0)$ ($=74.5 (\Omega cm)^{-1}$,  for sample  SFMO(N1))
 is about 50 times lower than the metallic conductivity of an SFMO single crystal at T=0.{\cite{tomioka}}

The upper curve in Fig. 1(a) shows $\sigma(T)$ of a sintered SFRO
sample that underwent a short heat treatment at 500$^o$C in
Ar5$\%$H$_2$. The maximum indicates mixed grain-boundary and
metallic conductivity. A similar behavior is seen in Fig. 2 of
Ref. {\onlinecite{kobayashi}} for an SFRO sample sintered in Ar
atmosphere. Prolonged heat treatment of our sample in air at 400$^o$C
restored inter-grain  tunneling and the $\sigma(T)$ plot straightened up (see lower curve
in Fig. 1(a)). The solid line represents Eq. (\ref{eq:FIT}) fitted
to the experimental data. Fig. 1(b) shows three more plots of
$\sigma(T)$ for SFRO samples, including one for a c.p. sample. The data for the c.p. sample exhibit  unusual behavior at high
temperatures and Eq. (\ref{eq:FIT}) could be fitted to this line
only up to 250 K.

Fig. 2 presents plots of $\sigma(T)$ for c.p.
samples of SCWO and SCRO that were not included in the
previous reports {\cite{SCWO,SCRO}} since at that time
they did not seem relevant for the main issues of those papers.

The three parameters T$_1$, T$_o$ and T$_1$/T$_o$ are plotted versus M$_i$ in Fig. 3(a)- (c). While no correlations are seen in Figs. 3(a)-(b), Fig. 3(c) exhibits two remarkable features:

1. The data of T$_1$/T$_o$ (the tunneling constant $\pi\chi w/2$) for SFMO fall between 2 and 3, irrespective of microstructure of the samples. Within the FIT model this corresponds to the remarkable linearity of $\sigma$(T). The independence from microstructure hints at the presence of an intrinsic insulating boundary layer through which tunneling occurs, with well-defined electronic (and magnetic) structure.

2. The data of T$_1$/T$_o$  for cold pressed samples ({\it {i.e.}} for bare boundaries) lie close to a straight line that extrapolates to zero  near M$_i=5$ which corresponds to the d-gap (see Table I).  The possibility that such a simple analytical function  fits the dependence of $\pi\chi w/2$ on M$_i$  for this set of half-metallic c.p. samples requires further experimental and theoretical support.

Our analysis shows that the quality of the tunneling barriers in inter-grain conductivity depends on  T$_o$. The higher T$_o$ relative to RT, the closer is inter-grain tunneling to elastic tunneling.   Table II and Fig. 3(b) show that, for only 3 samples out of 19,
T$_o \ > 1000$ K, {\it i.e.} for two polycrystalline samples of SFMO (in Fig. 3(b) the two symbols coincide) and for one c.p. sample of SFRO.
 Note that for SFMO(N1) the ratio $\sigma(RT)/\sigma(0)$ is only 1.25.
 The wide spread of the FIT parameters implies  a broad range of interactions governing  magneto-resistance (magneto-conductance). Results in Ref. {\onlinecite{JMMM}} show that for sintered SFMO samples the magneto-conductance is much higher than that for a c.p. sample at all temperatures. This requires a more systematic investigation.

\pagebreak


\pagebreak

\begin{table}
  \centering
  \caption{The five Sr$_2$BB'O$_6$ half-metals, their ionic configurations, nominal saturation magnetization per formula unit (M$_i$) and Curie temperature (T$_c$ (K))  }

\vspace{1cm}

\begin{tabular}{|c|c|c|c|}
  \hline
  \hline
  Sr$_2$BB'O$_6$& Ionic configuration
  & M$_i$ ($\mu_B$/f.u) & T$_c$ (K) \\
  \hline
  \hline
Sr$_2$FeMoO$_6$& Fe$^{3+}$ (3d$^5$)Mo$^{5+}$(4d$^1$)&  4  & 420\\
(SFMO)&&&\\
 \hline
Sr$_2$FeReO$_6$& Fe$^{3+}$ (3d$^5$)Re$^{5+}$(5d$^2$)&  3  & 400\\
(SFRO)&&&\\
 \hline
Sr$_2$CrMoO$_6$& Cr$^{3+}$ (3d$^3$)Mo$^{5+}$(4d$^1$)&  2  & 450\\

(SCMO)&&&\\
\hline
Sr$_2$CrWO$_6$ {\cite{SCWO}} &  Cr$^{3+}$(3d$^3$)W$^{5+}$
(5d$^1$) &  2 &  390\\
(SCWO)&&&\\
\hline
Sr$_2$CrReO$_6$& Cr$^{3+}$(3d$^3$)Re$^{5+}$(5d$^2$)& 1& 635\\
(SCRO&&&\\
\hline
\hline

\end{tabular}\\

\end{table}

\pagebreak

\begin{table}
  \centering
  \caption{Fitting parameters for Eq. (\ref{eq:FIT}), for various Sr$_2$BB'O$_6$
  polycrystalline samples. }

\vspace{1cm}

\begin{tabular}{|c|c|c|c|c|c|c|}
  \hline
  \hline
  Sample& Data source & M$_i$        & $\sigma(0)$ & $T_1$(K)& $T_o$ (K)& $T_1/T_o$ \\

        &             & ($\mu_B$/f.u)&($\Omega$cm$)^{-1}$ &         &           &           \\
  \hline
  \hline
 SFMO(N1)&Fig. 2 in Ref. {\onlinecite{efield}}& 4&74.5 &5171&2478&2.09\\
 \hline
 SFMO(r)&Fig.4 in Ref. {\onlinecite{remarkable}} &4&33.8 &6989&2551&2.74\\
 \hline
 SFMO(c.p.)&Fig. 2 in Ref. {\onlinecite{remarkable}}&4&0.72 &1586&582&2.73\\
 \hline
 \hline
SFRO (Ox)&Fig. 1(a) &3&2.50 &396&494&0.80\\
 \hline
 SFRO (S1)&Fig. 1(b) &3&3.88 &1117&703&1.59\\
 \hline
 SFRO (S2)&Fig. 1(b) &3&13.0 &276&397&0.70\\
 \hline
 SFRO (c.p.)&Fig. 1(b) &3&0.19 &8111&1329&6.10\\
 \hline
 \hline
SCMO(A)&Fig. 2(a) in Ref. {\onlinecite{remarkable}}  &2&1.02 &712&263&2.71\\
\hline
SCMO(C)& Fig. 2(a) in Ref. {\onlinecite{remarkable}} &2&2.48 &455&117&3.89\\
\hline
SCMO(E)& Fig. 2(a) in Ref. {\onlinecite{remarkable}}  &2&6.1 &390&130&3.00\\
 \hline
SCMO(A+Ox)&Fig. 2(b) in Ref. {\onlinecite{remarkable}}   &2&4.6$\times$10$^{-4}$ &1104&207&5.33\\
\hline
SCMO(D)&Fig. 2(b) in Ref. {\onlinecite{remarkable}}   &2&0.31 &1405&228&6.16\\
\hline
SCMO(B1)&Fig. 3 in Ref. {\onlinecite{remarkable}}   &2& 6.3$\times10^{-4}$& 1430&180&7.94\\
\hline
SCMO(B2)&Fig. 3 in Ref. {\onlinecite{remarkable}}   &2& 5.6$\times10^{-4}$& 1177&152&7.74\\

 \hline
 \hline
SCWO& Fig. 6(a) in Ref.{\onlinecite{SCWO}} &2&0.045 &260&118&2.20\\
\hline
SCWO& Fig. 6(a) in Ref.{\onlinecite{SCWO}} &2&0.030& 461&213&2.16\\
\hline
SCWO (c.p.)& Fig. 2 &2&2$\times 10^{-6}$ &1850&200&9.25\\
\hline
\hline
SCRO (c.p.)& Fig. 2&1&1.2$\times 10^{-4}$ &4071&383&10.63\\
\hline \hline
LSFMO (c.p.)&Fig. 4 in Ref. {\onlinecite{remarkable}} &3.5&0.47 &941&298&3.16\\
\hline \hline

\end{tabular}\\

 r - reduced, c.p. - cold-pressed, Ox - oxidized

\end{table}

\pagebreak

%
%

Figure Captions

\begin{figure}[h]
\begin{center}
\includegraphics[width=15cm]{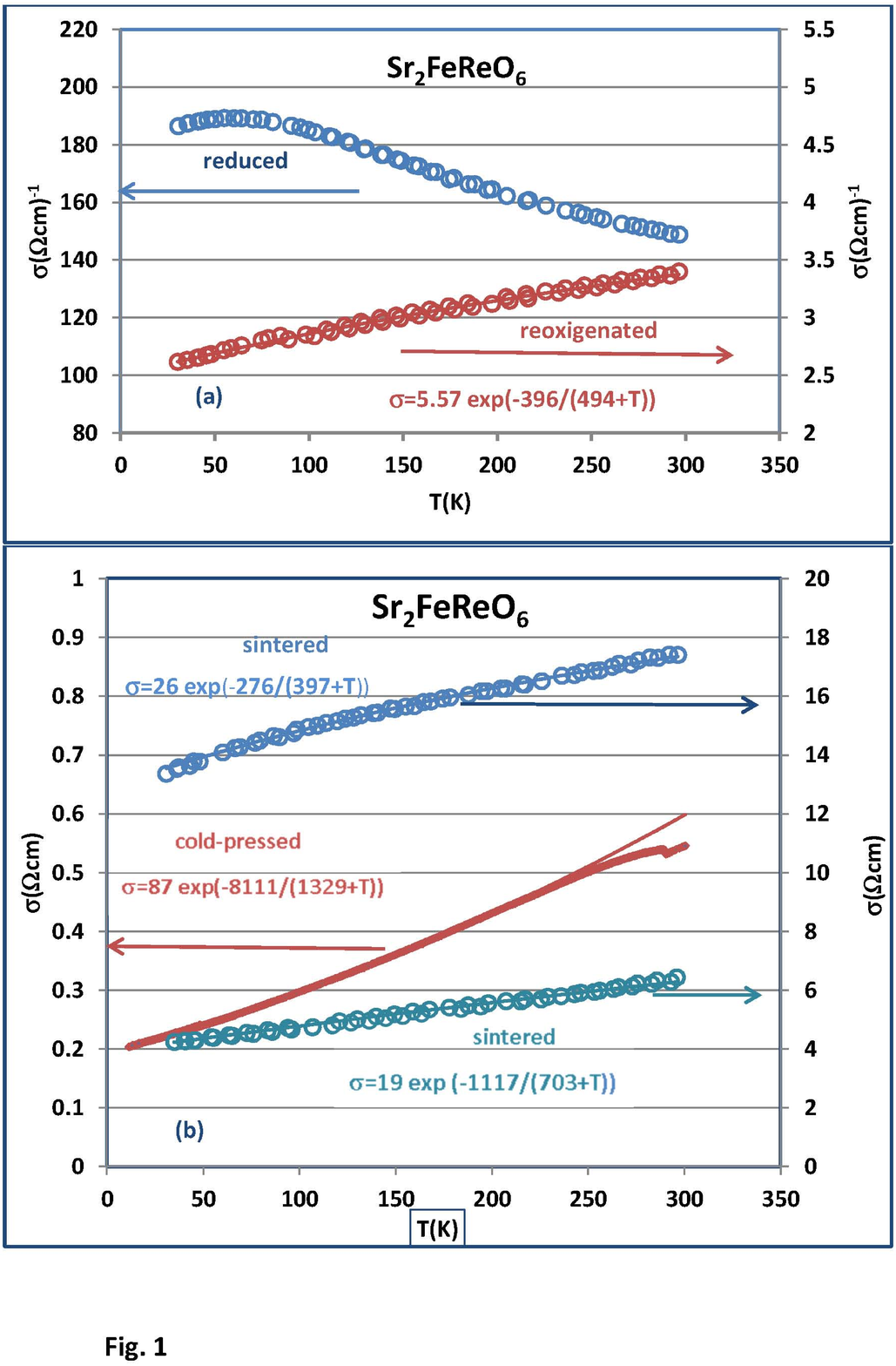}
\caption{Conductivity versus temperature of (a) a sintered sample of Sr$_2$FeReO$_6$   heat treated at 500$^o$ C in a reducing atmosphere (upper curve)  and later reoxygenated at 400$^o$C (lower curve), and (b)  two additional sintered samples and one cold pressed sample.  Solid lines in (a) and (b) represent Eq. 1 fitted to  experimental data.   } \label{1}
\end{center}
\end{figure}

\begin{figure}[h]
\begin{center}
\includegraphics[width=15cm]{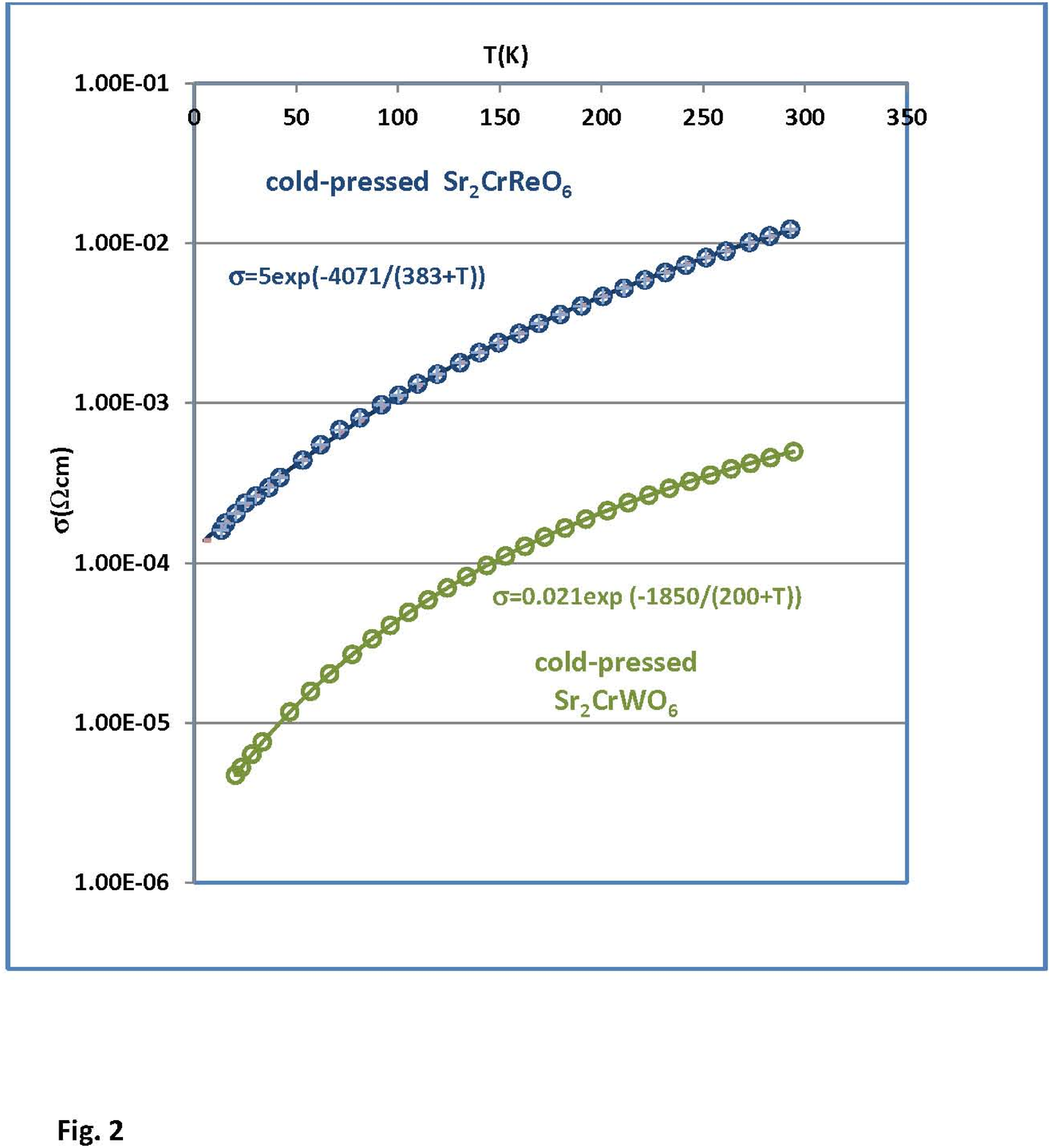}
\caption {Conductivity versus T of cold-pressed samples of Sr$_2$CrWO$_6$ (lower curve) and Sr$_2$CrReO$_6$ (upper curve). Solid lines represent Eq. 1 fitted to experimental data.} \label{2}
\end{center}
\end{figure}

\begin{figure}[h]
\begin{center}
\includegraphics[width=15cm]{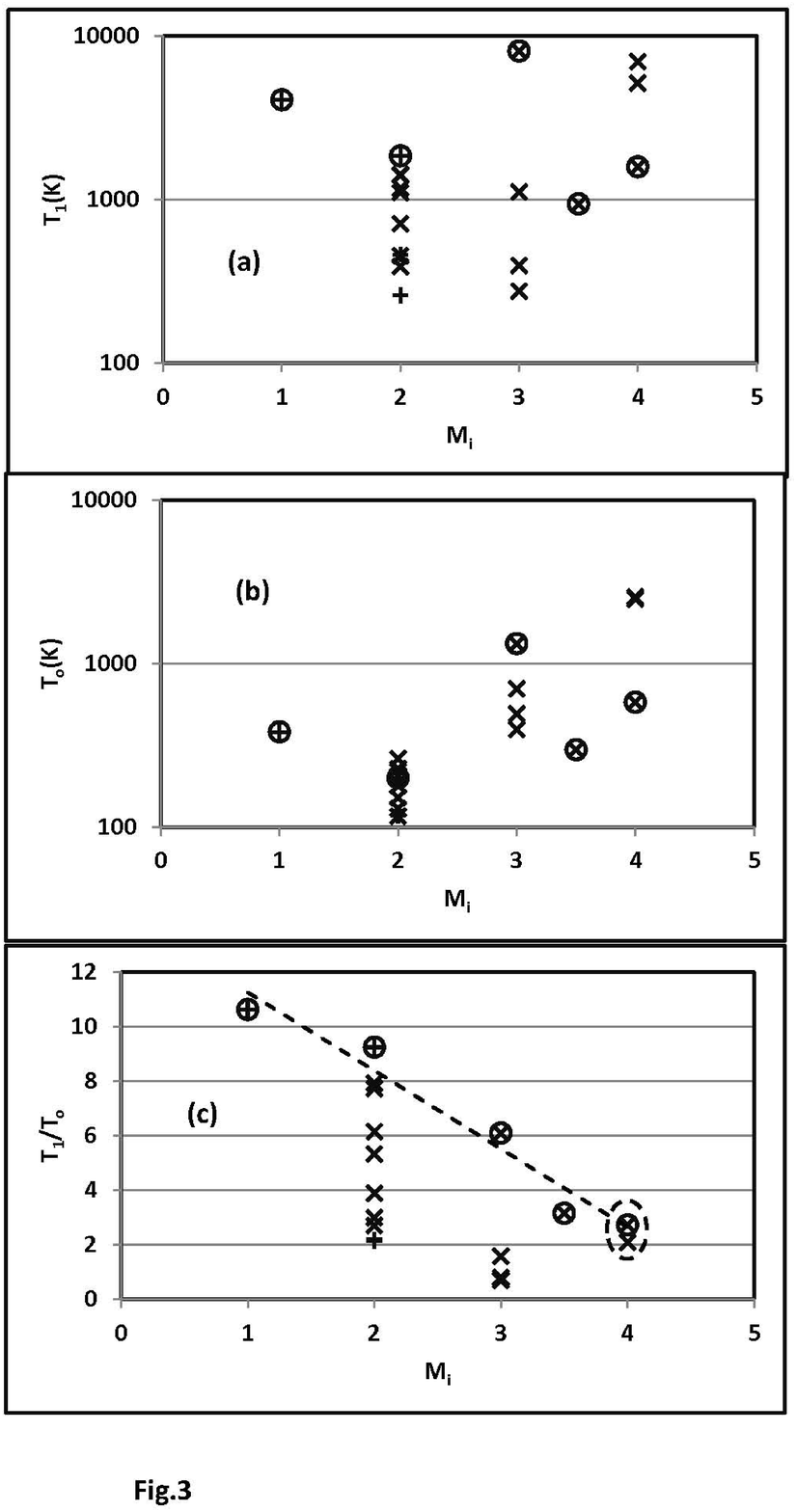}
\caption{Fitting parameters T$_1$, T$_o$ and  the tunneling constant - T$_1$/T$_o$ as function of M$_i$, the nominal saturation magnetization per formula unit. For M$_i$ =2 the symbol $\times$ represents samples of SCMO and + the samples of SCWO.  Encircled symbols represent data for cold pressed samples. Note that for all SFMO samples, T$_1$/T$_o$ falls between 2 and 2.75 (within the range of linear $\sigma$(T)). The values of T$_1$/T$_o$ for the cold pressed samples increase almost linearly with M$_i$.}
\end{center}

\end{figure}

\end{document}